# Origins of leakage currents on electrolyte-gated graphene field-effect transistors

A. Svetlova[1, 2], D. Kireev[1, 3, 4], D. Mayer[1], A. Offenhäusser[1, 2]

1 Institute of Biological Information Processing (IBI-3), Forschungszentrum Jülich GmbH, Wilhelm-Johnen Straße, Jülich, 52425 Germany

2 RWTH Aachen University, Aachen, 52062 Germany

ABSTRACT. Graphene field-effect transistors are widely used for development of biosensors. However, certain fundamental questions about details of their functioning are not fully understood yet. One of these questions is the presence of gate/leakage currents in the electrolyte-gated configuration. Here, we report our observations considering causes of this phenomena on chemical vapor deposition (CVD) grown graphene. We observed that gate currents reflect currents that occur on the transistor surface similarly to a working electrode – counter electrode pair currents in an electrochemical cell. Gate currents are capacitive when the graphene channel is doped by holes and Faradaic when it is doped by electrons in field-effect measurements. The Faradaic current is attributed to a reduction of oxygen dissolved in the aqueous solution and its magnitude increases with each measurement. We employed cyclic voltammetry with a redox probe $Fc(MeOH)_2$ to characterize changes of the graphene structure that are responsible for this activation. Collectively, our results reveal that through the course of catalytic oxygen reduction on the transistor's surface more defects appear.

Each year, graphene-based electrical devices are gaining more spotlight in a field of biological and biochemical applications, including sensing of toxic gases,[1] sensing of biomolecules,[2] cell action potentials[3, 4] and many more. For detection of solution-phase analytes, the graphene surface is commonly immersed in an aqueous solution. While graphene devices successfully operate in different configurations, field-effect transistors (FETs) gain the most from the graphene unique electrical properties.[5] Immersion of a gate electrode that controls an applied electric field in an analyte solution is a specific FET configuration that can be called "liquid gated", "electrolyte gated" or "electrochemically gated" FET. However, this configuration also bestows new effects that are usually not expected in a traditional back-gated or dielectric-gated FET devices, such as quite low but always existing leakage/gate currents ($I_g$). As mentioned in a recent review,[5] the exact nature of gate currents on an electrolyte-gated graphene transistors is not clear yet, as well as any possible impact that this effect would have on a measurement. In this letter we present our observations considering $I_g$ phenomena on CVD graphene FETs and suggest methods of characterization.

The GFET arrays used here, were fabricated as described in our previous works[3] and a schematic is shown in Figure S1. In order to shine more light on the gate current leakage phenomena, same GFETs were tested in two electrical schemes. The first one (see Figure 1A) corresponds to a standard electrolyte-gated field-effect transistor configuration with source and drain electrodes that contact both edges of graphene sheet. Second configuration (see Figure 1B) represents a standard 3-electrode scheme for cyclic voltammetry with a reference electrode (RE) and a counter electrode (CE). Out of two GFET feedlines, only one is connected to a potentiostat, making a graphene surface a working electrode (WE). Out-of-plane currents through a working electrode-liquid interface are measured.

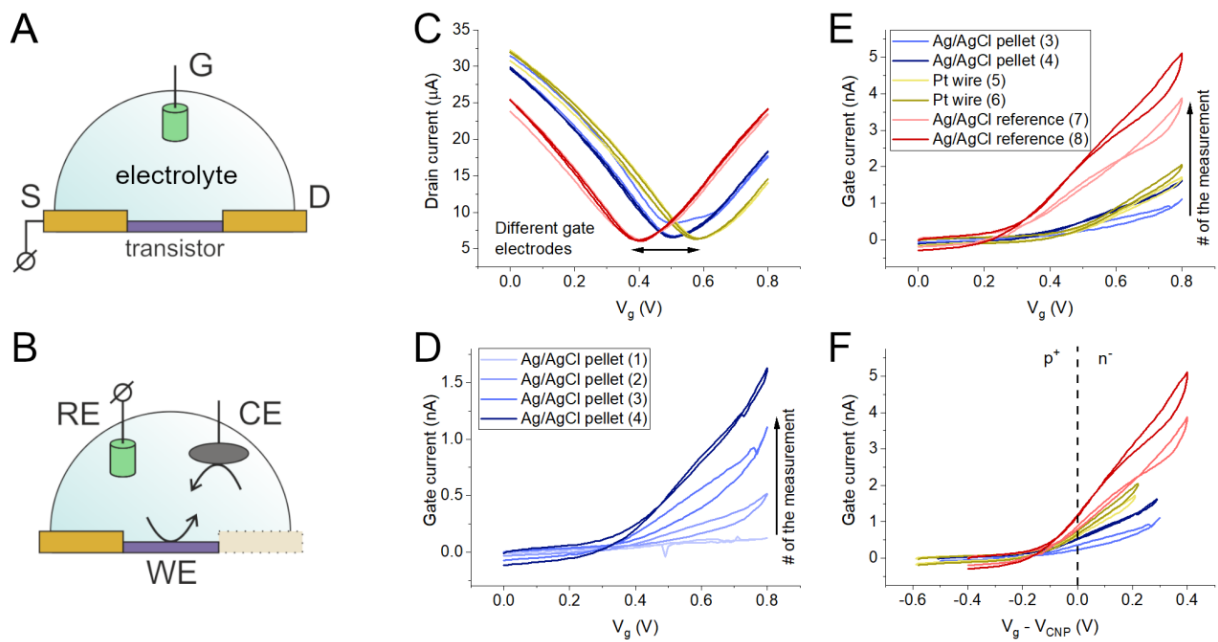

**Figure 1.** A. Schematics of a liquid-gated FET configuration. S – source feedline, D – drain feedline, G – gate electrode. B. Schematics of a 3-electrode electrochemical configuration. WE - working electrode, CE - counter electrode, RE - reference electrode. C. *I-V* scans with different gate electrodes. $V_g$ – applied gate voltage. D. Gate current increase during initial scans. E. Gate currents recorded with different gate electrodes. F. Gate currents normalized to $V_{CNP}$. Figures C-F share a color legend: blue lines – Ag/AgCl pellet gate, yellow lines – Pt wire gate, red lines – Ag/AgCl Dri-Ref gate. Number in brackets states a serial number of a measurement.

When just fabricated, the GFETs require one to three voltage sweep scans before they will feature stable and reproducible current-voltage characteristics (*I-V*), see Figures 1D and S2. This instability during first scans can be partially attributed to a current-induced removal of photoresist residues[6] or hydrocarbons[7] adsorbed from exposure in ambient air. This is a commonly discussed personally yet rarely mentioned in scientific literature feature of electrolyte gated GFETs. In a stable *I-V* regime all working transistors have excellent conductivity with maximum transconductance in the range of 20-60 µS (see Figure S3). The charge neutrality points ($V_{CNP}$) of transistors within a group with the same geometrical parameters showed a good accordance (see Figure S4). While performing *I-V* sweeps, gate leakage currents were simultaneously measured. In the case of electrolyte-gated FETs, these currents pass through the gate electrode-electrolyte interface. The recorded $I_g$ that we observed during *I-V* scans followed a similar profile on all devices: while insignificant at the small positive potentials, the magnitude of $I_g$ increases rapidly once a certain gate potential is applied. The maximum $I_g$ in a recording corresponds to a maximum set up gate potential as these signals did not show an apparent plateau. $I_g$ currents feature different maximum values on different transistors, which also increase with each consecutive scan (an illustrative example of a signal profile is presented on a Figure 1D). There is no dependence of $I_g$ maximum value on the overall number of measurements from an array of transistors made with a gate electrode. After 1-4 sequential scans on one transistor the maximum reached value increased from ~100 pA to several nA and occasionally to >10 nA. This increase is individual for each transistor. When proceeding to the characterization of another transistor, the pattern repeated: gate current started from below 1 nA and increased with a total number of scans – thus we conclude that this characteristic is a property of the transistor and not the gate electrode. To confirm the hypothesis, we performed same transistor's *I-V* characterization with different gate electrodes: Ag/AgCl pellet, Pt wire, and Ag/AgCl Dri-Ref electrode. These results are presented on Figure 1C-F. Although the choice of the gate electrode did not affect the magnitude of the drain-source current, it affected the position of $V_{CNP}$ which is a feature that was previously reported[8]. The $V_{CNP}$ shift results from a different potential drop over the gate-electrolyte interface for different gate materials. In this way each gate delivers a different "effective potential" on a surface of a transistor at an identically set $V_g$. Measured gate currents followed the same profile on all tested gate electrodes with a visible potential offset between them while a maximum current value continued to increase with each additional measurement (see Figure 1E). When the gate potential was normalized to $V_{CNP}$, it is clear that the observed current was greater in an electron-doping branch of a transistor functioning with no regards to the material of the gate electrode (see Figure 1F).

The characteristic features of gate currents signify that they are caused not by events on a gate-electrolyte interface, but by events on the transistor surface. The surface of CVD graphene transistors has an ability to act as a non-polarizable electrode in a solution.[9] This means that electrons can transfer between solution and a transistor surface as an out-of-plane currents at appropriate potentials. The capacitive character of $I_g$ as an interfacial charging current is visible at potentials up to 0.2 V negative in relation to $V_{CNP}$, Figure 1F. Once graphene-electrolyte potential-dependent Faradaic currents occur in an electron doping area, a gate electrode is forced to act as a counter electrode for this cell, similarly to a standard electrochemical cell. Currents observed on a gate-liquid interface have oxidative character ($I_g > 0$), which means that the transistor-liquid interface is involved in a reduction reaction. A possible reaction that can take place in an electrolyte solution is an oxygen reduction reaction (ORR). Although it was known for a long time that an electrochemical doping by $O_2/H_2O$ redox pair occurs on graphene surface[10, 11], this does not explain a sequential increase of the current in a consecutive series of *I-V* scans as seen on Figure 1D-E, which indicates a surface activation towards this process.

To elucidate the impact that the presence of leakage currents has on graphene properties, we characterized the same transistors intermittently, by cyclic voltammetry (CV) with redox probe 1,1′-Ferrocenedimethanol (Fc(MeOH)$_2$) diluted in 100 mM KCl followed by timetrace current sampling (*I-t*) at $V_{CNP}$. The full characterization pathway is shown on Figure 2A and an expanded description is provided in SI. An example of a time recording is shown on Figure 2B and gate current profiles for this transistor in a sweeping mode before (black) and after (red) the timetrace recording are presented on Figure 2C.

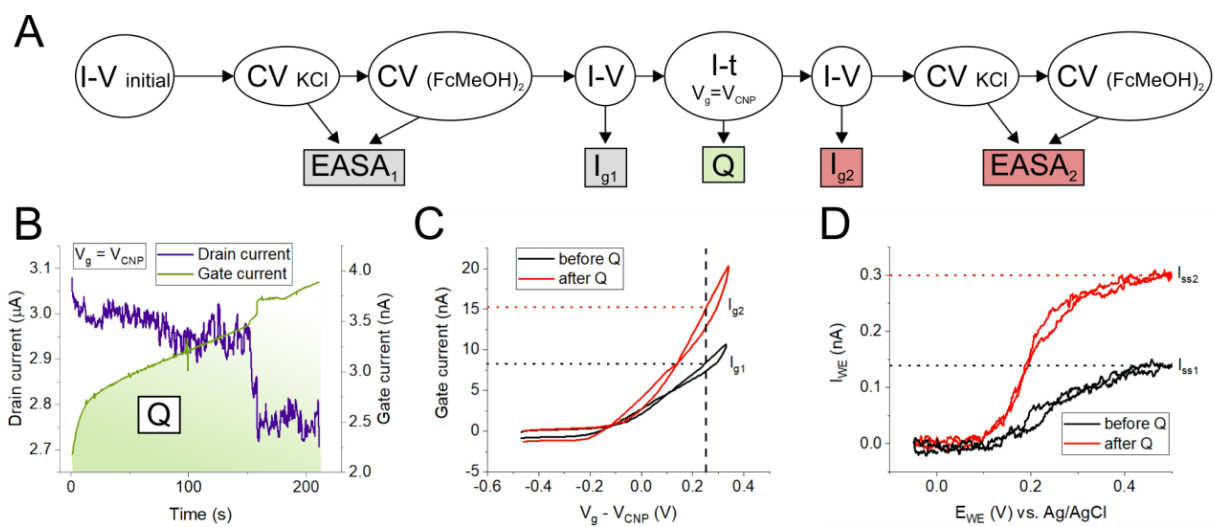

**Figure 2.** A. Characterization pathway. B. *I-t* recording of a transistor. Left axis, purple – drain current. Right axis, green – gate current. *Q* - total charge passed through a gate. C. $V_{CNP}$-normalized gate current during a voltage sweep before and after *I-t*. D. First and second background-subtracted CV recordings in 600 µM Fc(MeOH)$_2$. Scan rate 10 mV/s.

When transistors were characterized by cyclic voltammetry for the second time, an increase in oxidation currents was visible. An example of background-subtracted oxidation currents of the redox probe for the same transistor before and after *I-t* recording is presented on Figure 2D. In both cases, no oxidation happened until the potential of the working electrode reached the oxidation potential for Fc(MeOH)$_2$, after which in diffusion-limited conditions an oxidation current reached a plateau, which is called steady-state current ($I_{ss}$) and is a typical CV signal form for a microelectrode.[12] We used $I_{ss1}$ and $I_{ss2}$ to estimate an electrode electrochemically active surface area (EASA) as described in SI.

It is known that ferrocene compounds can show a residual adsorption on graphene.[13] This adsorption was estimated to be ~1% of a maximum surface coverage by ferrocene molecules.[14] We observed no residues of Fc(MeOH)$_2$ oxidation signal in a background KCl electrolyte scans (see Figure S5A), although it may be attributed to a low magnitude of a signal generated by low number of molecules on a microelectrode which did not exceed a background noise. After a CV measurement in Fc(MeOH)$_2$ solution, a $V_{CNP}$ shift and $g_m$ decrease were observed (see Figures S3, S4, S6A,C) which is similar to the one observed after adsorption of nucleotides on graphene surface.[15-17] Despite that, the character of the $I_g$ signal remained the same as before contamination: low values of the current in p-branch and growth of $I_g$ with an increasing $V_g$ in n-branch (see Figures S6B,D). CV measurements demonstrated a diffusion control of the probe oxidation at the used scan rate. Preliminarily, we can say that Fc(MeOH)$_2$ demonstrated good potential to estimate the EASA of CVD graphene surfaces since its presence didn't severely interfere with dynamics of the process behind $I_g$.

Overall changes in $I_g$ and EASA for all tested transistors are shown in Figure 3. We picked a value of $I_g$ at $V_g = V_{CNP} + 0.25$ V that the transistor reaches during an $I$-$V$ forward scan (see vertical line on Figure 2C) as a characteristic for these changes as currents in that area showed visible stepwise increase during the initial characterization. The absolute increase in this $I_g$ value between a scan taken directly before $I$-$t$ and a scan taken directly after $I$-$t$ ($I_{g1}$ and $I_{g2}$ on Figure 2C) in relation to a total charge $Q$ that passed through a transistor during $I$-$t$ (green area on Figure 2B) for all transistors are plotted in Figure 3A. Two populations of transistors are presented. The first (black markers) were not characterized by CV and were never in contact with Fc(MeOH)$_2$ solution.

The second (red markers) was exposed to the Fc(MeOH)$_2$ probe during CV. Nonetheless, both have linear correlation between $I_g$ increase and $Q$ with a similar slope and an offset of 1.5 nA between them. However, despite the increase in recorded values of $I_{ss}$ on all tested transistors (see an example on Figure 2D), we have found no correlation between $Q$ and EASA. The typical increase in $I_{ss}$ ranges from 30% up to 350%. On Figure 3B, the changes in calculated EASA are presented with a geometrical area (GA) of each transistor as well.

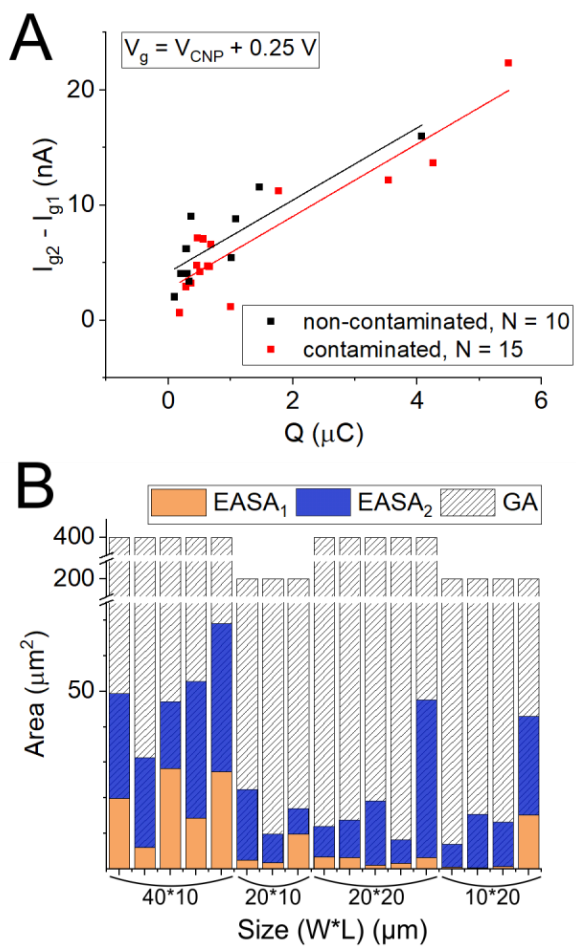

**Figure 3.** A. Dependence of gate current increase on total passed charge Q. Slopes of line fits: 3.1±0.6 nA/µS for transistors not contaminated with Fc(MeOH)$_2$ (black markers) and 3.1±0.4

nA/μS for transistors contaminated with traces of Fc(MeOH)$_2$ (red markers). B. Increase of electrochemically active area, EASA. GA – geometrical area.

Additional insights in the origins of the phenomena were provided by comparison of measurements in a regular and an oxygen-depleted electrolytes (see SI for experimental details). Results are shown on Figure 4 and S7. As one can see from Figure 4A, the GFETs exhibited almost identical performance in both regular, and oxygen-depleted electrolytes, with only a minor change in the CNP location. Notably, however, is the drastic decrease in the gate currents: both in a Faradaic region in a scanning *I-V* and in a timetrace *I-t* modes which links the process magnitude to a presence of dissolved oxygen in a solution.

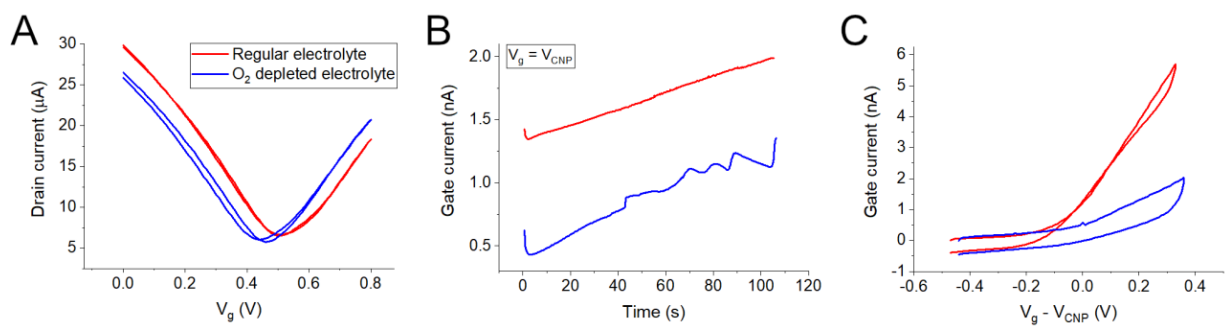

**Figure 4.** FET recordings in a regular electrolyte (red) and an oxygen-depleted electrolyte (blue). A. *I-V* scan. B. Gate current during *I-t* timetrace recording. C. Gate current in *I-V* scan.

As it was discussed, gate currents reflect events on a transistor surface, possibly an oxygen reduction reaction. Looking in detail on the ORR on graphitic surfaces, we can see that it can follow either a four-electron or a two-electron pathway with different intermediates depending on pH, physical properties of solution, and graphene doping.[18-20] In an acidic environments it leans toward a two-electron pathway that was shown to generate hydrogen peroxide H$_2$O$_2$, and oxygen-containing radicals[21] that can readily bind[22] to a graphene lattice. Structural doping of the graphene lattice by heteroatoms,[23, 24] oxygen-containing functional groups[25-28] and vacancy defects[29] enhances the catalytic performance of graphitic materials in the ORR. A commonly suggested mechanism of the reduction requires binding of an oxygen-containing intermediate to a carbon atom in a vicinity of a defect with a subsequent bond cleavage and various defects lower the energy barrier penalty of the transition state.[18] A CVD graphene after wet transfer is abundant in defects such as cracks, holes, carbon clusters of folds, and wrinkles.[30, 31] These clusters and photoresist residues that remain on a graphene surface[32] provide a source of carbon that can be oxidized during an initial *I-V* scans creating oxygenated defect sites. During our fabrication, shaping of the graphene sheet was done by oxygen plasma that creates oxygen-containing groups on graphene edges that were in contact with an analyte solution thus adding to the list of possible initial catalytic reaction centres. Incomplete cleavage of a C-O intermediate or generation of new oxygenated groups on a surface by peroxides would expand or activate the defect sites area[33], opening a higher surface area for an ORR (see Figure 5).

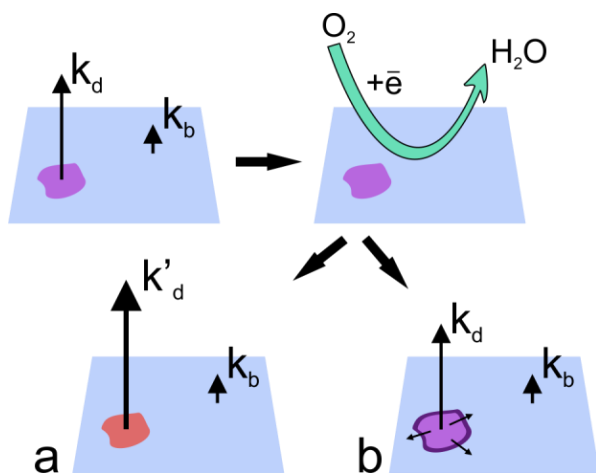

**Figure 5.** Suggested mechanism for gate current increase. Blue area is a graphene basal plane, purple colored area is a defect site, red colored area is an activated defect.

The purpose of CV measurements with Fc(MeOH)$_2$ was to elucidate structural changes that may accompany the gate current increase. Being an outer-sphere redox probe, Fc(MeOH)$_2$ molecules exchange electrons with a surface during oxidation in a density of states-dependent manner.[34] As expected from a low-defect electrode from a graphitic material that has density of states much lower than metals,[34, 35] the recorded oxidative currents presented on Figure 2B have relatively low magnitude in both cases. An increase of the plateau of the oxidation current as well as a shift of a halfwave potential $E_{1/2}$ to lower overpotentials can be attributed to an activation of the surface by higher number of defects.[36, 37] It is believed that defect sites have higher exchange rate ($k_d$ on Figure 5) than a basal surface of graphitic lattice ($k_b$ on Figure 5).[38] In this case a surface of a working electrode is considered electrochemically heterogeneous as it can be presented as two different electrode materials. In a case of a microelectrode and slow scan rates, only domains of material with faster electrode kinetics determine the overall signal as they consume redox species before the neighbour domains of slower kinetics material.[39] An increase in observed steady-state current can be attributed to two instances. The first one, Figure 5a, corresponds to an increase of the heterogeneous electron transfer rate from defects ($k'_d$) as a result of further oxidation of atoms within a defect ($k'_d > k_d$). The second one, Figure 5b, corresponds to a straightforward expansion of defects and increase of their surface area while the exchange rate remains the same.

Measurements of EASA increase after $I$-$t$ experiment in an O$_2$-depleted electrolyte may shed some light on EASA and $I_g$ link. Samples, presented on Figure S8 were exposed to Fc(MeOH)$_2$ solution. Nevertheless, on all transistors $I_g$ in a sweeping and a sampling modes are lower in oxygen-depleted electrolyte than in a regular electrolyte, similarly to non-contaminated samples (see Figure S8A-C). In a timetrace recording in an oxygen-depleted electrolyte $I_g$ values of these transistors are semi-stable for a duration of time with a fast boost that takes off at the end of the measurement (see Figure S8D-F). After these *I-t* measurements, two transistors had lower increase in $I_{ss}$ than transistors of the same geometrical parameters that are characterized in a regular electrolyte. One transistors' $I_{ss}$ did not increase after *I-t* in an $O_2$-depleted solution at all, Figure S8E-I. The $I_g$ value of this transistor in a sampling mode had shorter period of a fast offset compared to the other two which is an expected link of events within our theory. As in case of electrochemical oxidation[33], ORR-induced oxygen doping of the layer may be history-dependent for each transistor that carries a unique pattern of defects on its surface. Thus, an activation of each transistor will have a unique profile.

In conclusion we believe this report provides a description of previously overlooked phenomena of gate currents that is essential for comprehensive understanding of electrolyte gated graphene field effect transistors. We studied the relation of gate currents to the transistor properties and an oxygen-involving process. We propose the use of cyclic voltammetry as a complementary method sensitive enough to characterize miniscule changes that this process invokes in graphene structure. To preserve CVD graphene integrity, an extensive positive biasing of electrolyte-gated GFETs should be avoided, or oxygen-depleted solutions should be used.

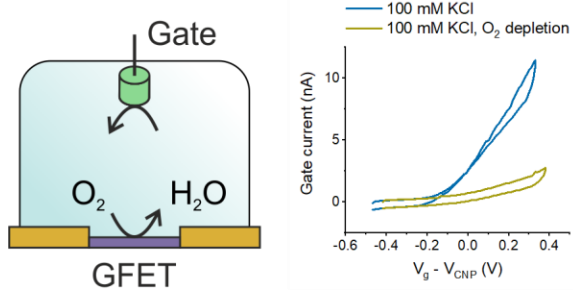

**Graphics for abstract**

ASSOCIATED CONTENT. **Supporting information.**

Explanation of two electrode schemes

Supplementary scheme 1

Experimental section

Supplementary figures S1-S8

AUTHOR INFORMATION

**Corresponding Authors**

A. Svetlova. Institute of Biological Information Processing (IBI-3), Forschungszentrum Jülich GmbH, Wilhelm-Johnen Straße, 52425 Jülich, Germany; orcid.org/0000-0002-9144-2474; a.svetlova@fz-juelich.de

D. Kireev. Department of Electrical and Computer Engineering, The University of Texas at Austin, Austin, Texas, 78758 USA; orcid.org/0000-0003-1499-5435; d.kireev@utexas.edu

**Present Address**

3 Department of Electrical and Computer Engineering, The University of Texas at Austin, Austin, Texas, 78758 USA

4 Microelectronics Research Center, The University of Texas, Austin, Texas, 78758 USA

**Author contributions**

A.S. conceived, performed and analyzed experiments with an input from D.K and D.M., D.K. fabricated GFET arrays, A.O. supervised the project, A.S., D.K. and D.M. wrote the manuscript, and all authors commented on it.

**Notes**

The authors declare no competing financial interest.

ABBREVIATIONS

CE, counter electrode; CV, cyclic voltammetry; CVD, chemical vapor deposition; EASA electrochemically active surface area; GA, geometrical area; GFET, graphene field-effect transistor; ORR, oxygen reduction reaction; RE, reference electrode; WE, working electrode